\documentclass[9pt,shortpaper,twoside,web]{ieeecolor}
\usepackage{generic}
\usepackage{cite}
\usepackage{amsmath,amssymb,amsfonts}
\usepackage{algorithmic}
\usepackage{graphicx}
\usepackage{textcomp}
\usepackage{multirow}
\usepackage{booktabs}
\def\BibTeX{{\rm B\kern-.05em{\sc i\kern-.025em b}\kern-.08em
    T\kern-.1667em\lower.7ex\hbox{E}\kern-.125emX}}
\begin{document}
\title{EVD Surgical Guidance with Retro-Reflective Tool Tracking and Spatial Reconstruction using Head-Mounted Augmented Reality Device}
\author{
    Haowei Li$^\dagger$, Wenqing Yan$^\mathsection$, Du Liu, Long Qian, Yuxing Yang, Yihao Liu, Zhe Zhao$^*$, Hui Ding, \\ Guangzhi Wang$^*$
    \thanks{
    This research was supported by National Key R\&D Program of China (2022YFC2405304), NSFC(U20A20389, U22A20355), Tsinghua University Clinical Medicine Development Fund (10001020508) and Medivis. Inc.
    }
    \thanks{Haowei Li$^\dagger$ and Wenqing Yan$^\mathsection$ have contributed equally to this work.
    }
    \thanks{H. Li, Y. Yang, G. Wang, and H. Ding are with the Department of Biomedical Engineering, Tsinghua University, China.
    }
    \thanks{W. Yan is with the School of Medicine, Tsinghua University, China.
    }
    \thanks{D. Liu is with the Department of Electronic Engineering, Tsinghua University, China.
    }
    \thanks{Z. Zhao is with the School of Clinical Medicine, Tshinghua University, China. Beijing Tsinghua Chang Gung Hospital, China.
    }
    \thanks{Y. Liu is with Laboratory for Computational Sensing and Robotics, Whiting School of Engineering, Johns Hopkins University, USA.
    }
    \thanks{Long Qian is with Medivis. Inc.}
    \thanks{* Corresponding authors.}
}

\maketitle

\begin{abstract}
Augmented Reality (AR) has been used to facilitate surgical guidance during External Ventricular Drain (EVD) surgery, reducing the risks of misplacement in manual operations.
During this procedure, the key challenge is accurately estimating the spatial relationship between pre-operative images and actual patient anatomy in AR environment.
This research proposes a novel framework utilizing Time of Flight (ToF) depth sensors integrated in commercially available AR Head Mounted Devices (HMD) for precise EVD surgical guidance. 
As previous studies have proven depth errors for ToF sensors, we first assessed their properties on AR-HMDs. 
Subsequently, a depth error model and patient-specific parameter identification method are introduced for accurate surface information. 
A tracking pipeline combining retro-reflective markers and point clouds is then proposed for accurate head tracking. 
The head surface is reconstructed using depth data for spatial registration, avoiding fixing tracking targets rigidly on the patient's skull.
Firstly, $7.580\pm 1.488 mm$ depth value error was revealed on human skin, indicating the significance of depth correction. 
Our results showed that the error was reduced by over $85\%$ using proposed depth correction method on head phantoms in different materials. 
Meanwhile, the head surface reconstructed with corrected depth data achieved sub-millimetre accuracy.
An experiment on sheep head revealed $0.79 mm$ reconstruction error.
Furthermore, a user study was conducted for the performance in simulated EVD surgery, where five surgeons performed nine k-wire injections on a head phantom with virtual guidance.
Results of this study revealed $2.09 \pm 0.16 mm$ translational accuracy and $2.97\pm 0.91$ degree  orientational accuracy.

{
}
\end{abstract}


\section{Introduction}
\label{sec:introduction}
External ventricular drain (EVD) is a typical neurosurgery to rapidly and effectively reduce intracranial hypertension, which is widely used in emergencies such as acute hydrocephalus, intra-ventricular hemorrhage, and increased intracranial pressure \cite{srinivasan2014history, kakarla2008safety}.
The conventional procedure of this surgery involves the manual insertion of a drainage tube through the skull, meninges, and brain parenchyma, relying on anatomical landmarks and pre-operative images.
Given that different patients have significant individual differences in anatomical structures, traditional manual EVD surgery faces significant challenges and uncertainties \cite{eftekhar2016app} and is highly dependent on the surgeon's experience \cite{pelargos2017utilizing}.
Previous studies have indicated that the accuracy of freehand drainage tube insertion during EVD surgery is only around 80\% \cite{huyette2008accuracy}.
Subsequently, repeated puncturations caused by incorrect insertion significantly increase surgical risks and complication probabilities \cite{mostofi2016surface,  wilson2013comparison, foreman2015external, park2011accuracy}.

Surgical navigation systems take advantage of pre-operative images, surgical path planning, and intra-operative tracking to provide spatial information during surgical procedures.
Therefore, multiple studies have utilized surgical navigation in EVD surgery for higher operation accuracy and lower surgical risks  \cite{gautschi2014non, shtaya2018image, fisher2021image}, and have proven their effectiveness. 
Traditional surgical guidance utilizes intra-operative CT scanning \cite{fiorella2014integrated}, optical tracking systems \cite{jakola2014three}, and electromagnetic tracking systems \cite{alazri2017placement} to provide spatial information during operation, and typically display guidance information on a two-dimensional screen away from the surgical area.
Such approaches result in complex setups, hand-eye coordination deficits, and distraction, preventing them from being widely adopted in daily practice \cite{wilson2013comparison, fried2016insertion}.

In contrast, Augmented Reality (AR) can overlay virtual guidance information on actual surgical areas, providing more intuitive spatial information without the distraction caused by extra screens. 
Among all the augmented reality display methods, Head-Mounted Devices (HMD) can provide free observation while minimizing interference with surgical areas.
Multiple studies have demonstrated the potential of AR-HMD in intra-operative neurosurgical navigation \cite{frantz2018augmenting,meola2017augmented,guha2017augmented, qi2021holographic}.
In EVD surgery, previous studies suggest that AR surgical guidance can enhance the  drainage tube insertion accuracy \cite{li2018wearable}, help medical students handle surgical techniques more quickly \cite{van2021effect}, and provide comparably higher efficiency than other surgical navigation methods \cite{yi2021virtual}.

\begin{figure*}[t!]
    \centering
        \includegraphics[width=2.0\columnwidth,page=1]{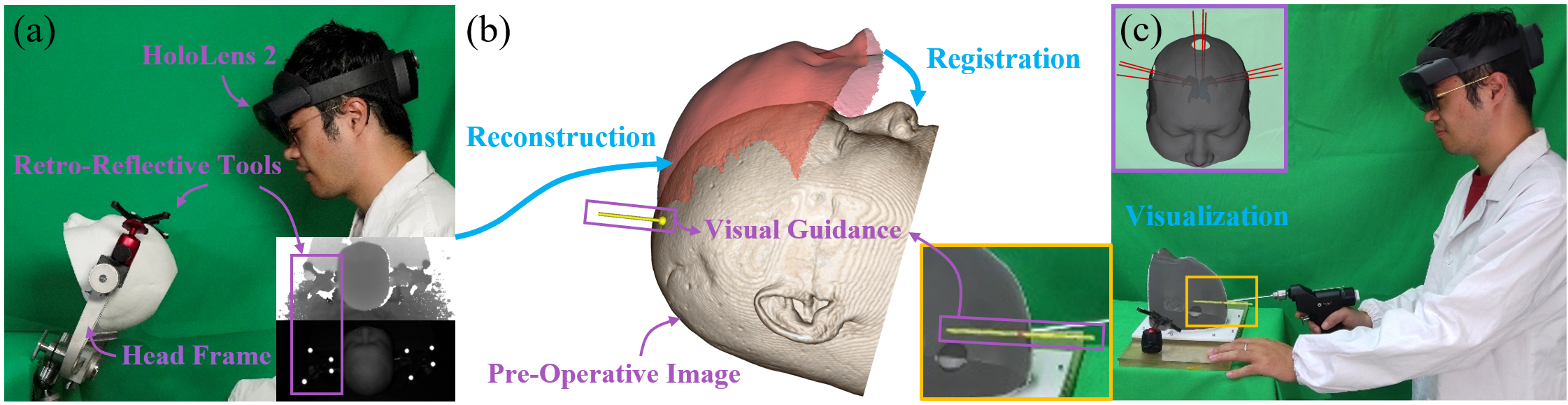}
    \caption{(a) The setup for accurate EVD surgical guidance utilized in the proposed framework. The retro-reflective tools are fixed on the headframe for tracking. (b) The sensor depth information is utilized to reconstruct the head surface and register pre-operative image with retro-reflective tool. (c) Retro-reflective tool tracking is used to provide visual guidance for drainage tube insertion. Here, we utilized k-wire injection to simulate EVD surgery.}
    \label{fig:TeaserFigure}
\end{figure*}

Acquiring precise spatial relationships between pre-operative images, the patient's head, and HMDs is one of the most critical problems in providing accurate AR navigation in EVD surgery.
Previous studies commonly use optical tracking targets for pose estimation.
In this approach, the targets are fixed rigidly and invasively on the patient's head during pre-operative imaging to obtain spatial relationships between images and the tracking tools  \cite{carbone2018proof}.
Alternatively, depth sensors on commercial AR-HMDs provide a non-invasive approach using three-dimensional surface information but are limited by the accuracy and efficiency of sensor depth value \cite{Gu2021DepthError,palumbo2022mixed}.
Commercially available surgical navigation systems typically employ retro-reflective tools and infrared cameras for tracking.
These tools, compositing of several retro-reflective markers with a unique spatial arrangement, are proven to be locatable under Time of Flight (ToF) depth sensors in AR-HMDs, providing more accurate and stable spatial information compared with visible light tracking methods \cite{kunz2020infrared, martin2023sttar}.
This approach is also highly efficient and robust to the change of environmental light \cite{HL2ResearchModel}.
Therefore, taking advantage of tool tracking and surface information, ToF depth sensors can potentially provide accurate spatial relationship acquisition methods featuring high efficiency, non-invasive, and high accuracy.

This work introduces a novel framework to enable accurate EVD surgical guidance using solely ToF depth sensors on commercially available head-mounted augmented reality devices.
This procedure combines retro-reflective tool tracking and surface information to ensure accuracy and effectiveness while reducing invasion.
Previous studies have illustrated a systematic error of depth value in on-device ToF depth sensors \cite{Gu2021DepthError}, which can also be found in different ToF sensors \cite{he2017depth} and varies according to the target material, depth values, and other factors \cite{tanaka2017material}.
These indicate that the ToF sensor would have different depth errors on different patients' heads. 
Therefore, the depth error of the ToF sensor on AR-HMD is first evaluated and patient-specifically corrected for the accurate head surface.
By utilizing a retro-reflective tool as the reference for accurate estimation of ToF sensor poses, corrected depth values can then be used to reconstruct dense and smooth head surfaces.
This is then used to register the spaces of the retro-reflective tool and pre-operative images.
During the intra-operative procedure, the tool is used to accurately track the patient's head, converting the pre-operative planned path to AR space, thus enabling in-situ guidance.

Compared to previous studies utilizing AR-HMD for in-situ surgical guidance, the main contribution of this work lies in the following:
1) We proposed a method for patient-specific ToF sensor depth value correction to acquire more accurate surface information, which takes advantage of pre-operative images and does not require complex setups such as fixing either the AR-HMD or the patient's head.
2) Utilizing the retro-reflective tool as a reference, we realized head surface reconstruction with multiple frames of ToF sensor depth data in sub-millimetre accuracy, reducing the sensor's data noise.
3) We proposed a pipeline combining retro-reflective tools and surface information for head tracking with high accuracy, high efficiency, and no extra invasion.

\section{Related Works}

Previous works have employed both external tracking systems \cite{de2019hand, barai2020outside} and built-in sensor resources in AR-HMD \cite{wang1990tracking, dho2021development} to realize pose estimation and in-situ display.
Comparatively, using built-in sensors for tracking can avoid both the integration of additional devices and the difficulties of positioning the tracking device in a cluttered operating room when avoiding view obstruction.
In the aspect of tracking methods, visible light markers \cite{mitsuno2017intraoperative,carbone2018proof,liebmann2019pedicle,qian2019aramis,lysenko2022use}, retro-reflective markers \cite{kunz2020infrared,van2021effect,gsaxner2021inside,martin2023sttar} and marker-less methods \cite{Gu2021DepthError,palumbo2022mixed} have been utilized for target tracking in surgical scenarios.
Among these methods, commercially available surgical navigation systems have widely adopted retro-reflective markers as they satisfy disinfection requirements and provide high accuracy during detection.

Unlike commercially available multi-view retro-reflective tool tracking systems, ToF depth sensors frequently implanted in AR-HMDs provide depth information and possibilities to distinguish each marker using infrared light reflectivity easily, thus enabling tool tracking with solely one sensor. 
Kunz et al., for the first time, proved the feasibility of using the ToF sensor integrated on AR-HMDs to track retro-reflective tools \cite{kunz2020infrared}.
Although depth data acquired from ToF sensors have relatively high error and instability compared with commercial multi-view systems, this approach still provided persuasive tracking accuracy ($0.76 mm$).
Alejandro et al.'s study extended this method to support retro-reflective tool definition, multi-tool tracking and pose filtering \cite{martin2023sttar}.
Their study proved retro-reflective tool tracking could provide higher accuracy than visible light approaches and enable accurate in-situ guidance.

Despite the capabilities for tool tracking, ToF depth sensors also provide massive surface information, enabling marker-less tracking with minimized invasion.
Palumbo et al. employed a single frame of point-cloud from the ToF depth sensor on AR-HMD to provide in-situ guidance for EVD surgery in the emergency room \cite{palumbo2022mixed}.
In an experiment including seven surgeons, this method provided $9.26\pm 3.27mm$ accuracy, $42\%$ higher than the control group with free-hand ($16.2\pm 3.48mm$).
Gu et al. utilized the same sensor to realize in-situ display for glenoid and explored the relationship between tracking accuracy and the occlusion of the target's surface \cite{Gu2021DepthError}.
They pointed out that a depth error of several millimetres existed on the ToF sensor implemented in AR-HMD, which varied according to the material and depth values.
This error contributed significantly to the displacement in surgical guidance.

Our work takes both advantages of retro-reflective tool tracking and surface information provided by on-device ToF depth sensors.
Compared with the studies enabling retro-reflective tool tracking with AR-HMD, on the one hand, our work validated and compensated the depth value error of ToF sensors on retro-reflective materials to achieve higher tracking accuracy.
On the other hand, by employing surface information from the ToF sensor, the proposed framework does not require fixing the tracking target on the patient's head during imaging, thus reducing invasion.
Compared with the marker-less approaches, we proposed a method to correct ToF sensor data patient-specifically to provide more accurate depth information.
Meanwhile, head surface reconstruction using depth images was realized to decrease noises in point clouds and improve registration accuracy.
Instead of solely point-cloud, the incorporated retro-reflective markers in the framework ensure the method's robustness to incidents such as patient movement.

            
\section{Methods}
\label{sec:Methods}
\begin{figure}[htpb]
    \centering
        \includegraphics[width=1.0\columnwidth,page=1]{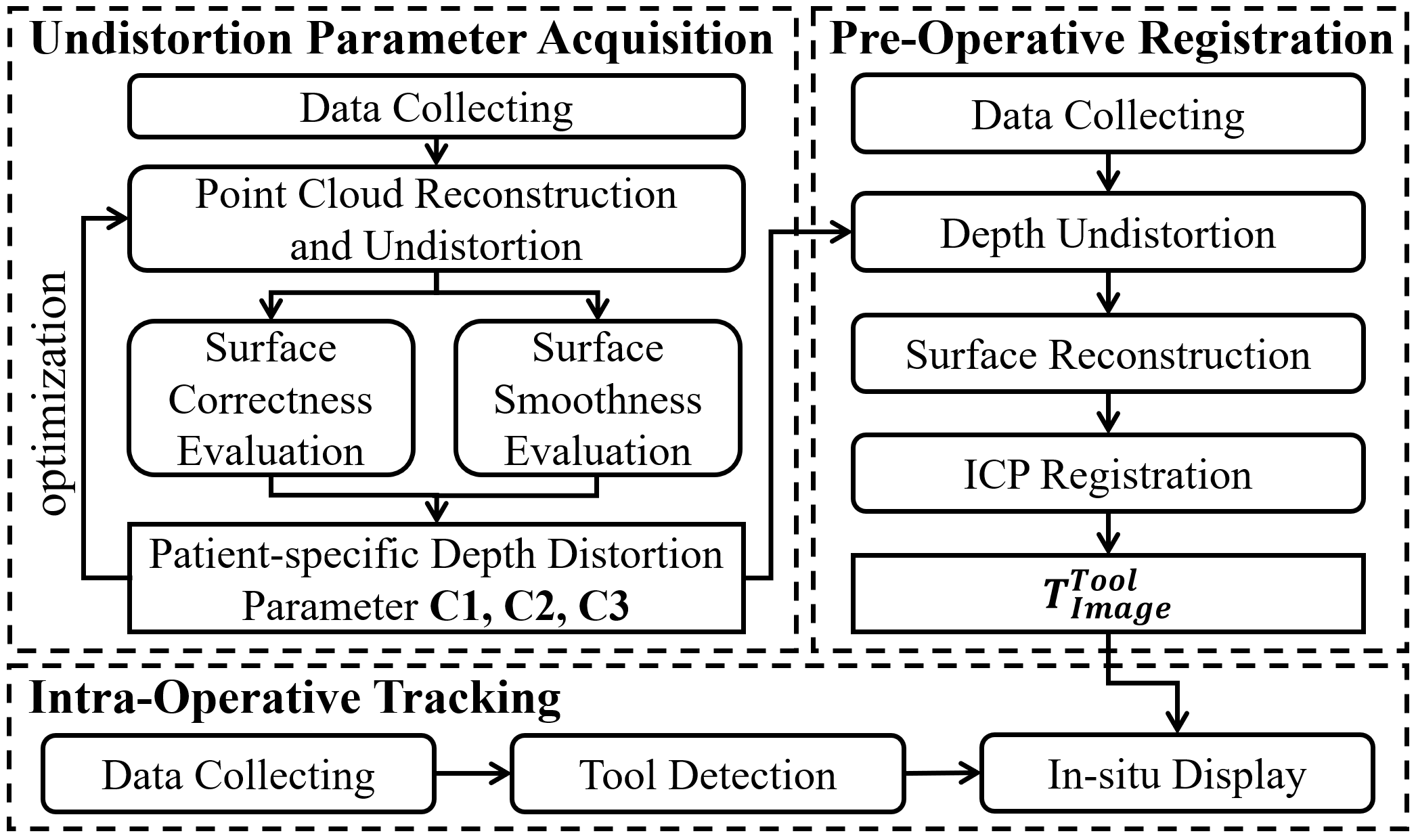}
    \caption{The proposed pipeline enabling accurate head tracking composites of three parts: 1) Multiple frames of AHAT sensor data are used to obtain patient-specific depth distortion parameters. 2) The undistorted point clouds are used to reconstruct the head surface and to register the pre-operative image with the retro-reflective tool rigidly fixed on the headframe. 3) The pose of the retro-reflective tool is extracted from dynamic sensor data for in-situ tracking.}
    \label{fig:SystemStructure}
\end{figure}

This paper presents a novel approach, employing retro-reflective tool tracking and point cloud information from ToF depth sensors in commercially available AR HMDs, to achieve accurate head tracking without additive invasion. 
Microsoft HoloLens 2 is adopted in our work due to the wide acceptance by previous studies, and Articulated HAnd Tracking (AHAT) mode ToF depth sensor is utilized for higher data frequency and broader Field of View (FoV). 
The structure of the proposed head tracking pipeline is shown in Fig. \ref{fig:SystemStructure}.
This section details the methods employed to accomplish accurate and non-invasive tracking, including ToF depth sensor correction, pre-operative registration, and intra-operative tracking.
A series of experiments and user studies are conducted to showcase the performance of the proposed method at different stages.

\subsection{Depth Error Correction}
\label{subsec:DepthErrorCorrection}
A patient-specific depth value correction method is proposed to provide more accurate surface information using the on-device ToF sensor.
We first conducted a series of experiments to discover the properties of the depth error in HoloLens 2 AHAT camera, on different materials, depths and rotations.
Retro-reflective material, traditional 3D printing materials and human head skin are included in these experiments. 
A depth error model and corresponding depth error correction method are proposed according to these properties.
As different patients have significant differences in depth error, the parameters in the depth error model need to be identified patient-specifically. 
Therefore, a parameter identification method is finally proposed, considering the corrected surface's correctness and smoothness.

\subsubsection{Depth Error on Retro-Reflective Material}
\begin{figure}[htpb]
    \centering
        \includegraphics[width=1.0\columnwidth,page=1]{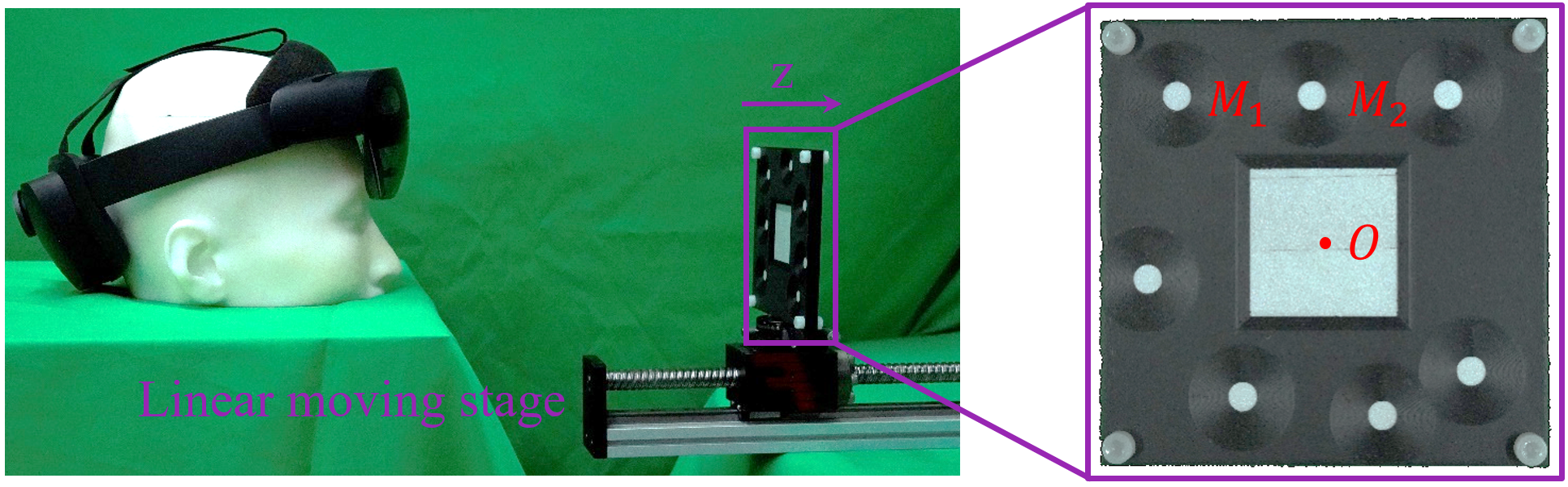}
    \caption{Testing structure to compare depth value provided by solving PnP and ToF sensor.}
    \label{fig:IRTapeDepthError}
\end{figure}

Though it is proven that by tracking retro-reflective tools with AHAT sensor, the system can provide accurate in-situ display \cite{martin2023sttar}, we still experimented if depth value is valid on retro-reflective materials to compare different materials fairly. 
The experiment structure is shown in Fig. \ref{fig:IRTapeDepthError} (a).
Retro-reflective tapes (3M 7610\footnote{Minnesota Mining and Manufacturing.}), typically used to manufacture planar retro-reflective tools, were affixed to a 3D printed structure where part of the tape was exposed for tracking and evaluation.
Seven circular areas ${M_i}$ were first used for pose estimation of the structure by solving Perspective n Point (PnP) problem, thus calculating the three-dimensional centre $O^C$.
This point was projected to the depth image as $O^I$, and the depth value at this point was acquired as $d^I$. 
During the experiment, the testing structure was moved at a constant speed by a linear moving stage, while HoloLens 2 was fixed rigidly in the space.
The depth information from the PnP method and depth sensor in time series can be annotated as $O^C(t)$ and $d^I(t)$, respectively.
As the linear moving stage moved at a constant speed, the three-dimensional centre position $O^C(t)$ should be linear with timestamp $t$.
Therefore, the trajectory was first fitted using linear fitting (denoted as $O^{'C}(t)$) to eliminate noise caused by tracking.
The depth error of the AHAT sensor on retro-reflective material was finally calculated as $\delta(t)=d^I(t)-||O^{'C}(t)||_2$.
Here $||\cdot||_2$ refers to the Euclidean length of a vector.
This experiment was repeated on different angles ($0^\circ$, $10^\circ$, $20^\circ$, $30^\circ$) between the testing structure and the sensor.

When the retro-reflective material faced the camera directly, a depth value error fluctuating between $0mm$ and $4mm$ was observed within $300mm$ to $700mm$ in depth.
The fluctuation is mainly composited of two parts: 
1) A component varied slowly with depth. 
2) A high frequency, random noise caused by the sensor and the truncation of AHAT sensor depth data \cite{HL2ResearchModel}.
To further acquire the relationship between systematic depth error and depth value, a spline smoothing method was utilized to reduce sensor noise. 
We denote the depth error along depth value as $f_{rf}(d)$, and calculate the weighted sum of the fitting error and the roughness of the fitted curve ($RSS(f_{rf})$):
\begin{equation}
    RSS(f_{rf})=p\sum_i (\delta(d)-f_{rf}(d))^2+\int_{min(d)}^{max(d)} D^2 f_{rf}(d)dd
    \label{euqation:ErrorSplineFit}
\end{equation}

Here, $\delta(d)$ denotes the depth error at each depth, $D^2f_{rf}$ refers to the $2^{nd}$ derivative of $f_{rf}$.
By minimizing this function with a small weight value $p$, we acquired a smooth curve indicating the systematic depth error at each depth $f_{rf}(d)$.
In this approach, a $4.25mm$ maximum error difference at different depths was revealed.
Moreover, the depth error was slightly changed when the orientations were different.
Among all depths, the maximal error difference between different angles was found to be $1.08 mm$ on average.
Though the results revealed that both depth and orientation contributed to depth error on retro-reflective, only depth is considered for correction due to several reasons:
1) The pixels to obtain depth value on spherical retro-reflective markers always face the ToF camera directly.
2) The orientation range enabling the detection of the planar retro-reflective tool was limited.
3) The influence of the orientation is considerably smaller than that of the depth on the error.

Therefore, for each pixel on retro-reflective material detected in the camera space $M(x,y,z)$, the real position of the marker $M'(x',y',z')$ can be calculated as:

\begin{equation}
    M^{'} = \frac{||M||_2-f_{rf}(||M||_2)}{||M||_2} \cdot M
    \label{euqation:markerCenterDistortionMethod}
\end{equation}

\subsubsection{Depth Error on Different materials}
\begin{figure}[htpb]
    \centering
        \includegraphics[width=1.0\columnwidth,page=1]{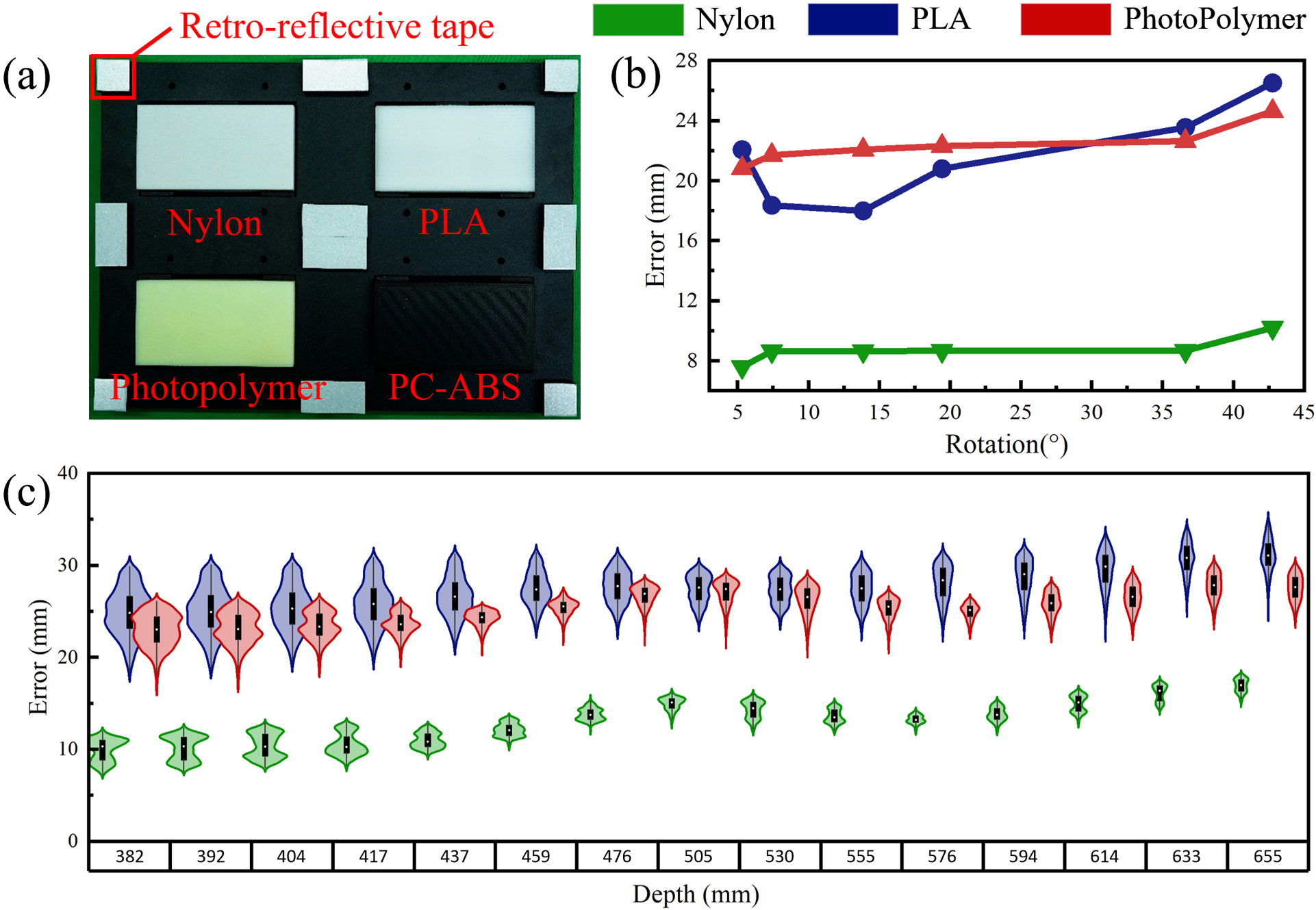}
    \caption{Depth errors of different materials under HoloLens 2 AHAT camera. (a) The testing structure holds all the materials and retro-reflective markers at the same plane. (b), (c) Depth errors of different materials at different angles and depths, PC-ABS is excluded here due to high depth instability and data missing rate.}
    \label{fig:PrintingMaterialError}
\end{figure}

The depth errors on different materials are then investigated using retro-reflective material as a reference.
To achieve this, a CNC-manufactured structure was used to hold all the testing materials and the retro-reflective material at a sample plane (see Fig. \ref{fig:PrintingMaterialError} (a)).
Four different materials commonly used in 3D printing, nylon, polylactic acid (PLA), photopolymer (PP) and polycarbonate-acrylonitrile butadiene styrene (PC-ABS), were included in this study. 
For each experimental group, the testing structure was positioned steadily at different depths and angles relative to the AHAT camera for 250 continuous data frames. 
Pixels on the retro-reflective material were first extracted from the sensor data, corrected using (\ref{euqation:markerCenterDistortionMethod}), and re-projected to 3D space.
These data points were then employed to fit the reference plane as $Ax+By+Cz+D=0$.
For any other point $X_i (x_{i},y_{i},z_{i})$ on the surface of testing materials, the actual spatial position $X_i{'} (x_{i}',y_{i}',z_{i}')$ was calculated by assuming that it lies on the reference plane: $Ax_{i}{'}+By_{i}{'}+Cz_{i}{'}+D=0$, $X_i'=(1-k)\cdot X_i$.
The depth error at each point was evaluated as $\delta_i = ||X_i||-||X_i'||=k||X_i||$.
This experiment was repeated at 15 different depths ranging from $382 mm$ to $655 mm$ and six different angles from $5^\circ$ to $45^\circ$

The results of this experiment are shown in Fig. \ref{fig:PrintingMaterialError} (b), \ref{fig:PrintingMaterialError} (c).
During the test, the depth value on PC-ABS was extremely in-stable ($std>5mm$), possibly due to the strong infrared light absorption caused by the black colour and the linear texture on the surface.
Therefore, it is omitted during data analysis.
As the distributions of depth errors at all conditions were proven non-normal, a Kruskal-Wallis test with a significance level of $\alpha = 0.05$, followed by a posterior Bonferroni test, was employed to uncover the disparities in depth errors under distinct conditions.
Our results suggested that the depth error distributions of different materials were significantly different, at all depths and angles.
The mean difference of the depth errors between PLA and Nylon reached $14.57 mm$ across all the depths.
Furthermore, those of the same material at different depths and angles were also significantly different.
The maximum change of the mean depth error at all the depths reached $6.10mm$, $5.29mm$ and $3.94mm$ for nylon, PLA and photopolymer, respectively.
The values for rotations reached $2.597mm$, $8.525mm$ and $3.814mm$.
In general, the depth error increased when depth values increased.
A non-decreasing tendency was revealed for depth errors on nylon and photopolymer when the rotation increased, which also existed in PLA when the rotation angle was larger than $15^\circ$. 
On the one hand, these results revealed a centimetre-level depth error existed on different materials, indicating the significance of depth correction when using AHAT depth information. 
This error significantly varied when the material was different. 
Therefore different materials should be treated differently during depth correction. 
On the other hand, as both depth and rotation contributed to the depth error, both factors should be considered during the correction. 

\subsubsection{Depth Error on Human Head}
To further prove the existence and diversity of the depth error in real situations, we conducted a study to reveal the AHAT sensor depth error on real humans' skin. 
In this study, a retro-reflective marker ($\Phi=10mm$) was attached to the forehead of the participant, while 500 frames of dynamic depth sensor data were collected within $180mm$ to $500mm$.
The marker in each frame was first extracted, and the depth values of the points on the marker were corrected.
After that, the points surrounding the marker were fitted with a cubic polynomial surface function.
The average distance from the points on the retro-reflective markers to the fitted surface was used to reveal the depth error of the skin near the marker.
Ten participants (5 males, 5 females, $M_{age} = 25.3$, $SD_{age} = 3.16$) were involved in this study.
As a result, an error of $7.580\pm 1.488mm$ was revealed in this study.
The substantial mean depth error on the human skin underscored the importance of depth correction for achieving accurate tracking using depth information. 
Furthermore, the variance in depth error across different individuals highlighted the necessity for patient-specific depth correction.

\begin{figure*}[t!]
    \centering
        \includegraphics[width=1.85\columnwidth,page=1]{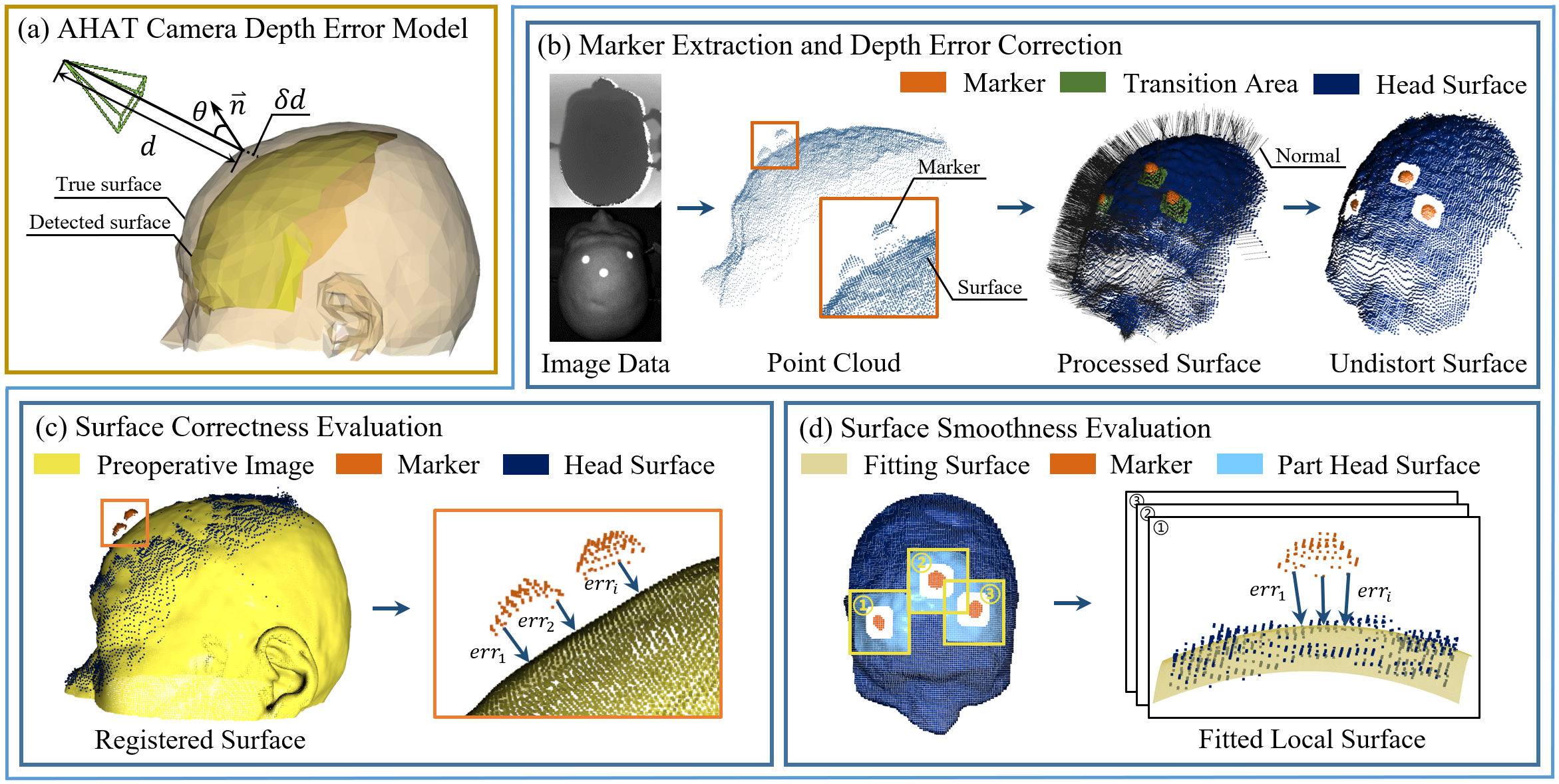}
    \caption{AHAT sensor depth error model and patient-specific parameter identification method. (a) HoloLens 2 AHAT sensor depth value distortion model. (b) Depth value correction procedure during patient-specific parameter identification. Multiple retro-reflective markers are stuck on the patient's head during this procedure. (c) Evaluate the correctness of the corrected surface using the pre-operative image. (d) Evaluate the smoothness of the corrected surface near each marker.}
    \label{fig:UndistortionMethod}
\end{figure*}
\subsubsection{ToF Sensor Depth Value Distortion Model}
Kenichiro et al.'s study explained each depth error component in the ToF sensor \cite{tanaka2017material}.
The band-pass characteristic of the target surface on the infrared light emitted by the ToF sensor creates a constant depth distortion, which may also explain the relatively low depth error on retro-reflective material. 
At the same time, the high-order harmonics components in the infrared light create depth-dependent errors. 
Moreover, Ying et al.'s work proved a linear relationship between depth error and depth value on ToF sensors \cite{he2017depth}. 

Considering the influence of depth value and rotation on the depth error, revealed by the experiments, and the conclusions provided by the previous studies, linear relationships between depth error and depth or rotation were considered.
The proposed depth error model is shown in Fig. \ref{fig:UndistortionMethod} (a).
For a given spatial point $P(x,y,z)$ in three-dimensional space, together with the norm of the surface at that point as $n(n_x,n_y,n_z)$, the distance from the point to camera $d=||P||_2$ and the angle of the local surface to the camera $\theta=\arccos(|n\cdot P|/||P||_2)$  were first calculated.
The depth error at this point can be calculated as follows:
\begin{equation}
    \delta d =C_1+C_2\cdot d+C_3\cdot \theta
    \label{equation:DistortionModel}
\end{equation}

The detected depth value can then be expressed as $d_r = d + \delta d$.
Consequently, given a detected surface point provided by AHAT sensor $\{P_i\}$, estimated norm $\{n_{P,i}\}$ and rotation angle $\{\theta_{P,i}\}$, we can calculate the true position of this point as:
\begin{equation}
    P_i^{'}=\frac{||P_i||_2-C_1-C_3\cdot\theta_{P,i}}{(1+C_2)||P_i||_2}\cdot P_i
    \label{equation:Undistortion}
\end{equation}

\subsubsection{Patient-Specific Parameter Identification}
To achieve accurate depth correction for each patient, the parameters $C_1, C_2, C_3$ presented in the model need to be identified patient-specifically.
Different from the traditional parameter identification method, the shape of the patient's skin is irregular, while the static relationship between the patient and the depth camera can not be promised.
Therefore, two evaluations of the corrected depth information utilizing pre-operative images and surface smoothness are proposed, which do not require the fixation of either the patient or AR-HMD.

\textbf{Data Acquisition and Pre-Processing:} 
To acquire the data for parameter identification, several retro-reflective markers are stuck on the patient's head as a reference.
Meanwhile, the HoloLens 2 is worn by the surgeon to record data at different depths and angles.
As shown in Fig. \ref{fig:UndistortionMethod} (b), an offset between the surface of the markers and the head exists when an inappropriate undistortion parameter is applied.
Moreover, a transitional area between markers and the head surface exists, where depth error is hard to predict. 
To provide effective surface information for parameter identification, each frame of sensor data $\{P_j\}$ is first separated to marker area, transitional area and head surface, where $j$ refers to the $j^{th}$ frame.
Marker area $\{P_{M,j}\}$ is first extracted using high infrared light reflectivity and is then dilated to generate the transitional area $\{P_{T,j}\}$. 
The remaining surface is regarded as targeting head surface $\{P_{H,j}\}$.
The angle of the surface to the camera at each point can be calculated as $\{\theta_{i,j}\}$ using the surface norm, where $i$ indicates the $i^{th}$ point in the frame. 
The transitional area $\{P_{T,j}\}$ is first removed during the depth correction.
The depth values in marker areas are first corrected using  (\ref{euqation:markerCenterDistortionMethod}), while those on the head surface are corrected using (\ref{equation:Undistortion}).
The points on the corrected marker surface and head surface are denoted as $P^{'}_{M,i,j}$ and $P^{'}_{H,i,j}$, respectively.

\textbf{Surface Correctness Evaluation:} 
The pre-operative images are first used to evaluate the correctness of the undistorted head surface.
As shown in Fig. \ref{fig:UndistortionMethod} (c), the undistorted head surface is registered with the surface extracted from pre-operative CT images.
The root mean square (RMS) value of the distance from marker points to the head surface is used to indicate the correctness of depth correction:
\begin{equation}
    Err_{C,j} = \sqrt{\sum_{i=1}^N \mathcal{D_C}(P^{'}_{M,i,j},\mathcal{I})^2/N}
\end{equation}

Here, $\mathcal{I}$ refers to the pre-operative head surface, in the form of a point cloud, $\mathcal{D_C}(P,\mathcal{G})$ refers to the minimum distance from point $P$ to point cloud $\mathcal{G}$, $N$ refers to point number on retro-reflective markers in each frame.

\textbf{Surface Smoothness Evaluation:} 
Given a group of appropriate model parameters, the surface should be continuous near the margins of the retro-reflective markers.
Therefore, an evaluation considering the smoothness of the corrected surface near each marker is used to further optimize model parameters and to address situations where pre-operative images are unavailable.
Firstly, a connected component analyzation is used to separate markers in each frame as $\{M_{j,k}\}$ (see Fig. \ref{fig:UndistortionMethod} (d)), where $k$ refers to the $k^{th}$ marker in each frame. 
For each marker $M_{j,k}$ with points $\{P^{'}_{M,i,j,k}\}$, the surrounding head surface is extracted as $\{P^{'}_{H,i,j,k}\}$. 
A cubic polynomial equation is used to fit these head points, providing a smooth expression of the surface around the markers, denoted as $\mathcal{S}_{j,k}$.
The RMS of the distance from $\{P^{'}_{M,i,j,k}\}$ to $\mathcal{S}_{j,k}$ across different frames and markers is used to evaluate the smoothness of the undistorted surface:
\begin{equation}
    Err_{S,j,k} = \sqrt{\sum_{i=1}^N \mathcal{D_S}(P_{M,i,j,k},\mathcal{S}_{j,k})^2/N}
\end{equation}

$\mathcal{D_S}(P,\mathcal{S})$ refers to the minimum distance of point $P$ and polynomial surface $\mathcal{S}$.

\textbf{Model parameter identification: }
Finally, the depth distortion model parameters for the targeting surface can be calculated by minimizing the sum of the cost functions to evaluate the surface correctness $Err_{C,j}$ and smoothness $Err_{S,j,k}$, over all the frames and markers:

\begin{equation}
    \begin{split}
    (C_{O,1},C_{O,2},C_{O,3})=\mathop{\mathrm{argmin}}\limits_{(C_1,C_2,C_3)\in \mathcal{R}^3}(\sum_{j=1}^J Err_{C,j}^2 
    \\
    + \sum_{(j,k)=(1,1)}^{(J,K)} Err_{S,j,k}^2 / K)
    \label{equation:MinimizeUndistortError}
    \end{split}
\end{equation}

Here $J$ refers to the frame number of AHAT sensor data for depth distortion model parameter identification, and $K$ refers to the number of markers stuck on the head surface. 
\subsection{Pre-Operative Registration}
A headframe is typically used in neurosurgery to keep the head steady during operation.
Instead of fixing any rigid tracking target on the patient's head, which would cause extra invasion and may commonly require additional imaging for tool-image registration, we place the retro-reflective tool on the head frame only during the intra-operative procedure and use surface information provided by ToF sensor for markerless registration.
Such an approach can also avoid modification of the existing surgical procedure.
A simulating setup of the proposed framework is shown in Fig. \ref{fig:UseCaseRealScene} (a).

To achieve accurate registration between pre-operative images and the retro-reflective tool, the head surface is first reconstructed in the space of the retro-reflective tool utilizing ToF depth information.
The surgeon first wears the HoloLens 2 to obtain continuous ToF sensor data targeting at patient's head at different depths and angles.
The spatial relationships between the retro-reflective markers are first constructed to generate the tool's coordinate.
By detecting the tool in each sensor frame as $T_{tool,i}^{AHAT}$, the sensor pose can be represented by $T_{AHAT,i}^{tool}$, providing an accurate estimation of sensor's poses.
After that, the depth value at each pixel is corrected utilizing (\ref{equation:Undistortion}) and identified patient-specific model parameters.
GPU-based TSDF reconstruction method and marching-cube surface extraction method are then implemented with depth data and camera poses to reduce the depth noise from the ToF sensor and acquire accurate, dense surface information.
Taichi computing structure \cite{TaichiOrigin} is used here to achieve high efficiency.
The head surface is then extracted from pre-operative images.
A point-to-plane ICP registration method \cite{ICPRegistration} is finally applied to register the pre-operative head surface and the reconstructed surface, providing spatial transformation from the coordinate of pre-operative to retro-reflective tool $T_{img}^{tool}$. 

\subsection{Intra-Operative Tracking}
During the surgical operation, the retro-reflective tool rigidly fixed on the headframe is used for head tracking. 
Compared to the methods solely using point cloud for tracking, this method can provide accurate tracking results with higher efficiency and ensure the robustness to the movement of both the patient and the surgeon. 
We implemented the same retro-reflective tool tracking method as the previous study \cite{martin2023sttar}.
Differently, the spatial position of each retro-reflective marker is corrected using (\ref{euqation:markerCenterDistortionMethod}) before pose estimation.
By continuously tracking the retro-reflective tool, the position of the preoperative image in HoloLens space can be calculated as:
\begin{equation}
    T_{img}^{world} = T_{rig}^{world}T_{AHAT}^{rig}T_{tool}^{AHAT}T_{img}^{tool}
    \label{equation:PoseCalculation}
\end{equation}

Here, $T_{img}^{tool}$ refers to the registration result of the pre-operative image and the reference tool, $T_{tool}^{AHAT}$ refers to the dynamic tracking result at the current frame, $T_{AHAT}^{rig}$ refers to the transform from AHAT sensor to rigNode of HoloLens 2 \cite{HL2ResearchModel}, and $T_{rig}^{world}$ refers to the pose of the rigNode in HoloLens 2 world space.
Both pre-operative images and planned surgical paths can be displayed in-situ using this transformation $T_{img}^{world}$.

\subsection{System Construction}
To streamline the proposed pipeline to enable accurate in-situ EVD guidance, a system combining AR-HMD and open-source medical imaging software 3D Slicer \cite{pieper20043d} is constructed.
This system does retro-reflective tool tracking \cite{keller2023hl2irtracking} and rendering on HoloLens 2 device, integrated as a D3D project, while surface reconstruction and registration are finished on 3D Slicer.
A network interface between HoloLens 2 device and 3D Slicer is implemented based on TCP and UDP protocol.
The AHAT sensor data and retro-reflective tool tracking results are streamed to the 3D Slicer for processing.
An ROI selection procedure is first implemented, utilizing a single frame of AHAT sensor data.
After that, multiple data frames are collected and corrected while the surface is reconstructed within the selected ROI.
As ICP registration requires a good initializing pose, the pre-operative image is manually aligned coarsely with the reconstructed surface in 3D Slicer, after which point cloud registration is used for accurate registration.
This registration result is then synchronized to the AR device.
While the on-device ToF depth sensor is used for dynamic tracking of the retro-reflective tool, pre-operative images and planned surgical routine can be overlaid in situ for visualization.

\section{Experiments and Results}
\label{sec:Results}
\subsection{Surface undistortion accuracy}
\begin{figure}[htpb]
    \centering
    \includegraphics[width=1.0\columnwidth]{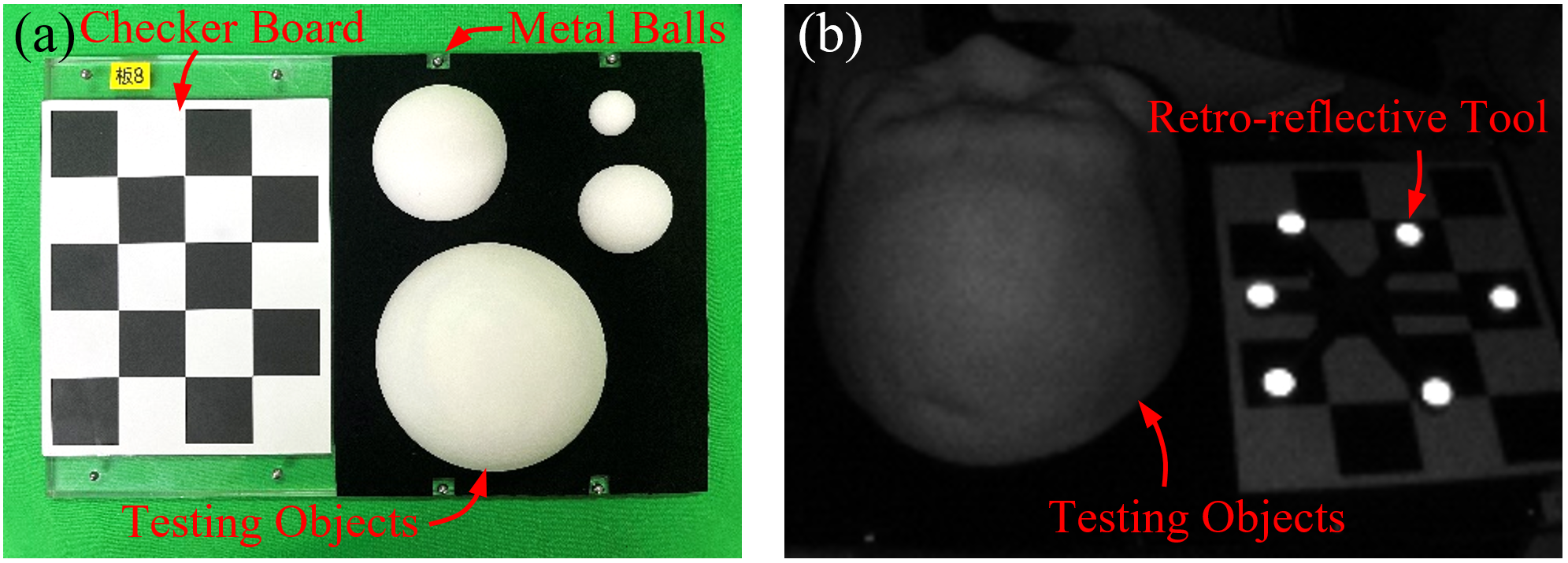}
    \caption{(a) Testing setup to evaluate the effectiveness of depth correction. (b) Testing setup for the accuracy of the reconstructed surface.}
    \label{figure:UndistortionAndReconstructionTestingStructure}
\end{figure}

We first conducted a series of experiments to evaluate the proposed ToF sensor depth value correction method.
The setup of this experiment is shown in Fig. \ref{figure:UndistortionAndReconstructionTestingStructure}(a). 

For each study group, a checkerboard and the target testing objects were rigidly attached to a flat acrylic board encircled by eight metal balls, which were visible in CT imaging. 
Before the experiment, the testing structure was CT scanned with $0.625mm$ spatial resolution. 
The surface of the target object $\{P_{O,i}^I\}$ and the centres of the metal balls in the volume $\{t_{B,i}^{I}\}$ were then extracted using a threshold.
After that, the metal balls were utilized to transform the object surface to the coordinate of the checkerboard.
The testing structure was kept static with a commercial IR tracking device (NDI\footnote{e.g., Polaris NDI, Northern Digital Incorporated. Ontario, Canada.}), which was used to extract the centres of the metal balls $\{t_{B,i}^{N}\}$ and the corners of the checkerboard $\{t_{C,i}^{N}\}$ using two retro-reflective tracking pointers.
The metal balls were then converted to the space of the checkerboard as $\{t_{B,i}^{C}\}$, utilizing singular value decomposition (SVD).
The transformation from imaging space to the checkerboard $T_{I}^{C}$ was calculated using the metal ball positions in both spaces.

\begin{figure}[htpb]
    \centering
    \includegraphics[width=1.0\columnwidth]{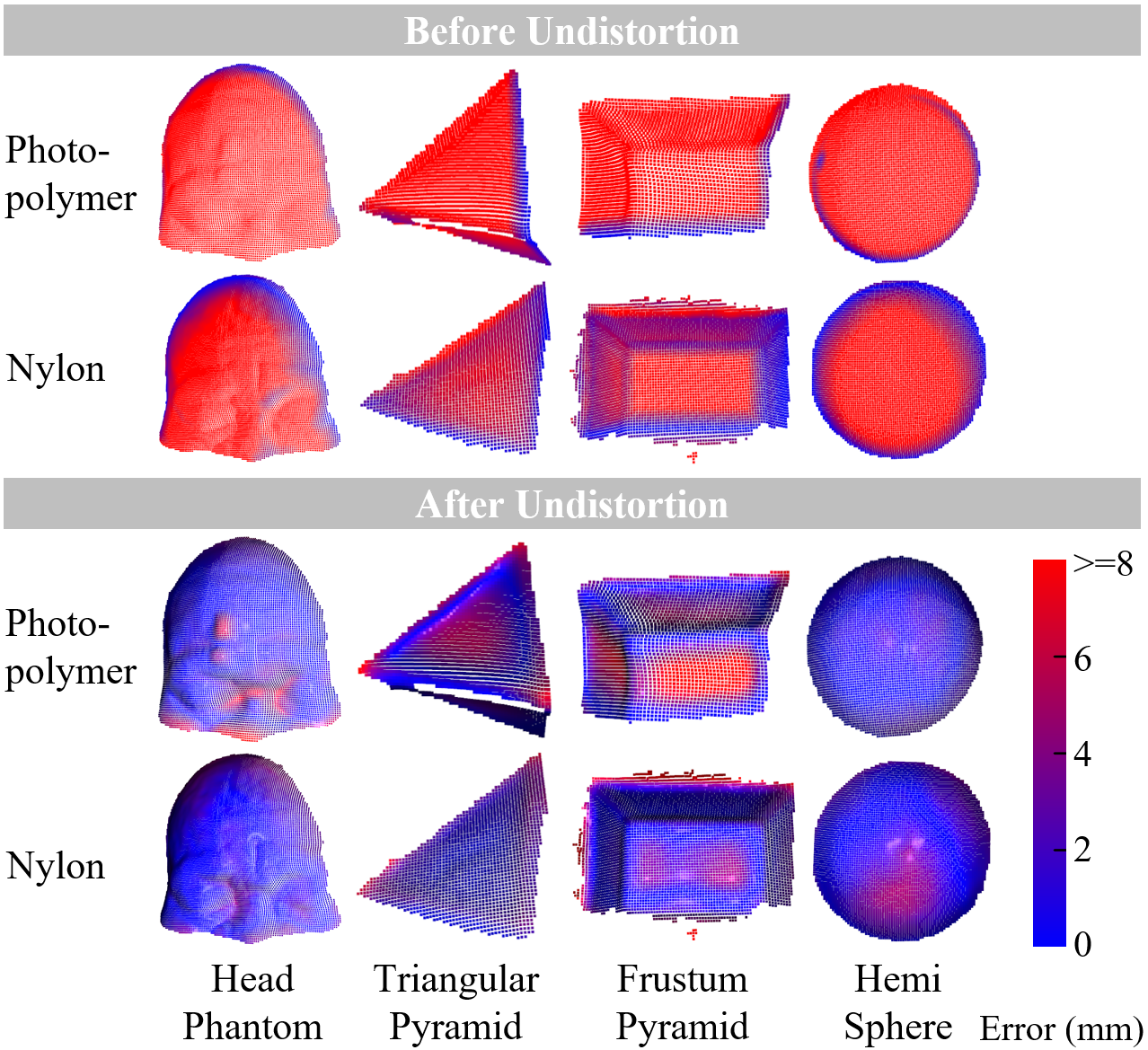}
    \caption{Error distributions of depth value accuracy before and after depth correction, tested on head phantoms and basic shapes in similar sizes of the head. This experiment was repeated on two different materials, photopolymer and nylon.
    }
    \label{figure:ErrorUndistortAccuracyHead}
\end{figure}

\begin{table*}[htpb]
\centering
\caption{Median value and IQR of depth error (mm) with and without depth undistortion.}
\begin{tabular}{ccccccccccccccccccc}
\toprule
\multirow{3}{*}{Material}     &    & \multicolumn{8}{c}{Before Undistortion}                                                                           &  & \multicolumn{8}{c}{After undistortion}                                                                            \\ \cmidrule{3-10} \cmidrule{12-19} 
                              &    & \multicolumn{2}{c}{Size 1} & \multicolumn{2}{c}{Size 2} & \multicolumn{2}{c}{Size 3} & \multicolumn{2}{c}{Size 4} &  & \multicolumn{2}{c}{Size 1} & \multicolumn{2}{c}{Size 2} & \multicolumn{2}{c}{Size 3} & \multicolumn{2}{c}{Size 4} \\
                              &    & Med          & IQR         & Med          & IQR         & Med          & IQR         & Med          & IQR         &  & Med          & IQR         & Med          & IQR         & Med          & IQR         & Med          & IQR         \\ \hline
\multirow{4}{*}{PP} & H  & 13.19        & 7.31        &              &             &              &             &              &             &  & 0.89         & 1.14        &              &             &              &             &              &             \\
                              & TP & 10.63        & 8.12        & 9.18         & 8.10        & 6.88         & 4.12        & 5.82         & 2.67        &  & 2.30         & 2.62        & 1.70         & 2.27        & 2.81         & 2.63        & 4.32         & 1.24        \\
                              & FP & 12.27        & 5.05        & 9.97         & 7.73        & 6.17         & 3.24        & 4.25         & 3.47        &  & 2.72         & 3.32        & 2.50         & 2.24        & 2.80         & 2.23        & 7.60         & 2.02        \\
                              & HS & 11.88        & 5.90        & 8.42         & 5.92        & 14.23        & 7.58        & 7.69         & 4.28        &  & 0.98         & 0.82        & 3.05         & 2.32        & 3.89         & 3.26        & 4.05         & 1.59        \\
                              & AVG& 11.59        & 6.36        & 9.19         & 7.25        & 9.09         & 4.98        & 5.92         & 3.47        &  & 2.00         & 2.25        & 2.42         & 2.28        & 3.17         & 2.71        & 5.32         & 1.62        \\
                              \cmidrule{2-19}
\multirow{4}{*}{Nylon}        & H  & 7.64         & 5.36        &              &             &              &             &              &             &  & 1.08         & 1.41        &              &             &              &             &              &             \\
                              & TP & 5.56         & 2.26        & 2.85         & 4.03        & 3.36         & 2.11        & 2.55         & 2.51        &  & 1.61         & 1.60        & 2.19         & 2.19        & 3.04         & 3.63        & 2.64         & 2.11        \\
                              & FP & 4.75         & 3.01        & 5.48         & 4.35        & 4.78         & 3.79        & 3.67         & 3.08        &  & 1.43         & 1.71        & 1.63         & 2.18        & 2.58         & 2.70        & 4.11         & 3.14        \\
                              & HS & 7.59         & 6.71        & 6.89         & 6.57        & 5.56         & 5.12        & 3.97         & 4.09        &  & 1.48         & 1.75        & 1.82         & 1.89        & 1.30         & 2.56        & 2.87         & 2.04        \\ 
                              & AVG& 5.97         & 3.99        & 5.07         & 4.98        & 4.57         & 3.67        & 3.40         & 3.23        &  & 1.51         & 1.69        & 1.88         & 2.09        & 2.31         & 2.96        & 3.21         & 2.43        \\
                              \bottomrule
\multicolumn{19}{p{500pt}}{Med: Median; PP: PhotoPolymer.}\\
\multicolumn{19}{p{500pt}}{H: Head phantom; TP: Triangular Pyramid; FP: Frustum Pyramid; HS: Hemi Sphere; AVG: Average value over TP, FP and HS.}

\end{tabular}
\label{table:ResultDeptherror}
\end{table*}

\begin{figure*}[htb]
    \centering
        \includegraphics[width=2.0\columnwidth,page=1]{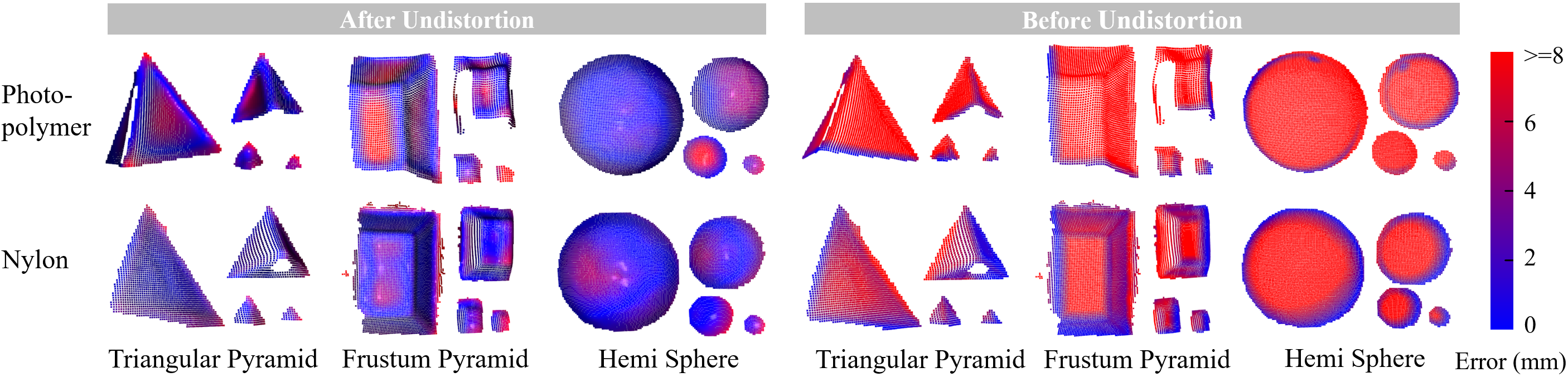}
    \caption{Error distributions of depth value accuracy before and after depth correction on models of four different sizes.}
    \label{figure:ErrorUndistortAccuracy}
\end{figure*}

Before assessing depth correction's effectiveness, 3D printed head phantoms were used to obtain material-specific depth correction parameters $C_1$, $C_2$, and $C_3$.
For each study group, a head phantom in the same material as the target objects was used for parameter identification.
Four retro-reflective markers, each with $10mm$ diameter, were stuck on the phantom's forehead, while 1000 frames of sensor data at random poses were collected for parameter identification utilizing (\ref{equation:MinimizeUndistortError}).
The testing structure was then kept facing the AHAT sensor directly and statically, at around $300mm$ in depth.
In this configuration, 300 continuous frames of data were collected for analysis. 
The poses of the checkerboard in the sensor space were first extracted utilizing infrared light reflectivity.
To eliminate noises from pose estimation, the average pose among all the frames $T_{C}^{S}$ was used to calculate the true position of the object surface in the sensor space as $P_{O,i}^S = T_{C}^{S}T_{I}^{C}P_{O,i}^I$.
After that, the mean value of the depth data across all the frames at each pixel was calculated to reduce the instability of AHAT sensor depth value, which was then undistorted with pre-calculated parameters using (\ref{equation:Undistortion}).
The distances from the detected surface to CT scanning result $\{P_{O,i}^S\}$ before and after depth correction were used to evaluate the effectiveness of the proposed method.

This experiment was first conducted on a head phantom and three models in common shapes, including a triangular pyramid, a frustum pyramid and a hemisphere.
These models were in similar sizes as the head phantom ($>100mm$ in the longest edge).
The error distributions in this experiment are recorded in Fig. \ref{figure:ErrorUndistortAccuracy}, and the quantitative results are recorded in Table. \ref{table:ResultDeptherror}.
As the distributions of the errors did not follow normal distributions, the median values and the interquartile ranges (IQR) were used for evaluation.
Moreover, the experiments were repeated on two different materials, photopolymer and nylon, to reveal the method's repeatability on different materials.

Generally, $0.89mm$ and $1.08mm$ depth errors remained on head phantoms made with photopolymer and nylon after depth correction.
These values were $93.25\%$ and $85.86\%$ smaller than those before depth correction.
The depth errors on the triangular pyramid, frustum pyramid and hemisphere were reduced by $78.36\%$, $77.83\%$ and $91.75\%$ on photopolymer, respectively, which were $71.04\%$, $69.89\%$ and $80.50\%$ for nylon.
Wilcoxon signed rank test revealed significant differences in depth error distributions for all the experimental groups before and after depth correction, with $p<0.001$.
The depth error reduction rates were higher on photopolymer, potentially due to higher initial errors on this material.
Regarding error distributions, more significant errors were found in areas with sharper edges, such as the corner of the eyes and the noses in the head phantom and the margin of the models.
When comparing models in different shapes, head phantoms had the least depth error after correction for both materials.
Though the triangular pyramid model and frustum pyramid model presented more significant errors than head phantoms, the error at the model's edge remained small.
However, the faces near the edge significantly contributed to the final error, which phenomenon was more evident on the photopolymer.

To better explore the proposed method's performance and limitations, the experiment was repeated on objects of four sizes ($>100mm$, $60mm$, $35mm$, $20mm$ on average in the longest edge).
To present the data analysis more clearly, we use size 1 to size 4 to denote the model sizes from the largest to the smallest.
Overall, despite the groups on the photopolymer frustum pyramid (size 4), nylon triangular pyramid (size 4) and nylon frustum pyramid (size 4), all the experiment groups revealed reductions in depth error after correction. 
Apart from the nylon triangular pyramid (size 3) and nylon frustum pyramid (size 4), all groups revealed smaller interquartile ranges (IQR) after correction.
The average depth error was reduced by $82.65\%$, $73.39\%$, $62.15\%$, and $-1.91\%$ for photopolymer from size 1 to size 4, respectively, which were $73.81\%$, $55.67\%$, $44.06\%$, $4.06\%$ for nylon. 
Average IQR reduction rate reached $62.70\%$, $67.94\%$, $41.44\%$, $52.73\%$ for photopolymer and $48.77\%$, $55.59\%$, $2.24\%$, $21.37\%$ for nylon.
A decreasing tendency of depth error reduction rate was revealed in both materials as the target model size decreased. 
On both photopolymer and nylon, the average median error over three standard models decreased when the models' size decreased before the depth data was corrected.
However, this phenomenon was reversed entirely after the depth values were corrected.

\subsection{Surface reconstruction accuracy}
An experiment is further conducted to prove the accuracy of the reconstructed surface using the proposed framework.
The setup of this experiment is shown in Fig. \ref{figure:UndistortionAndReconstructionTestingStructure} (b).
The target object was placed on an acrylic board, rigidly fixed with a retro-reflective tool with six planar IR markers for camera poses.
To ensure that every group of experiments was performed at similar depths and viewpoints, the testing structure was placed on a rotation platform, rotating at a constant speed. 
At the same time, the HoloLens 2 was fixed rigidly for data acquisition.
The rotation platform was placed at around $300mm$ to $500mm$ from the ToF sensor during the experiment.
Before the experiment, the material-specific depth distortion model parameters were acquired in the same way as the last experiment.
For each testing group, 2000 continuous frames of ToF sensor data were collected.
The target surfaces were reconstructed after depth correction using the retro-reflective tool as a reference.
A TSDF volume with $1mm$ voxel distance was used for each group.
Subsequently, the object surface obtained by CT scanning and the reconstructed surface was registered. 
The distance from the reconstructed surface to the pre-operative surface at each point was used to evaluate the reconstruction accuracy.

\begin{figure}[htpb]
    \centering
    \includegraphics[width=1.0\columnwidth]{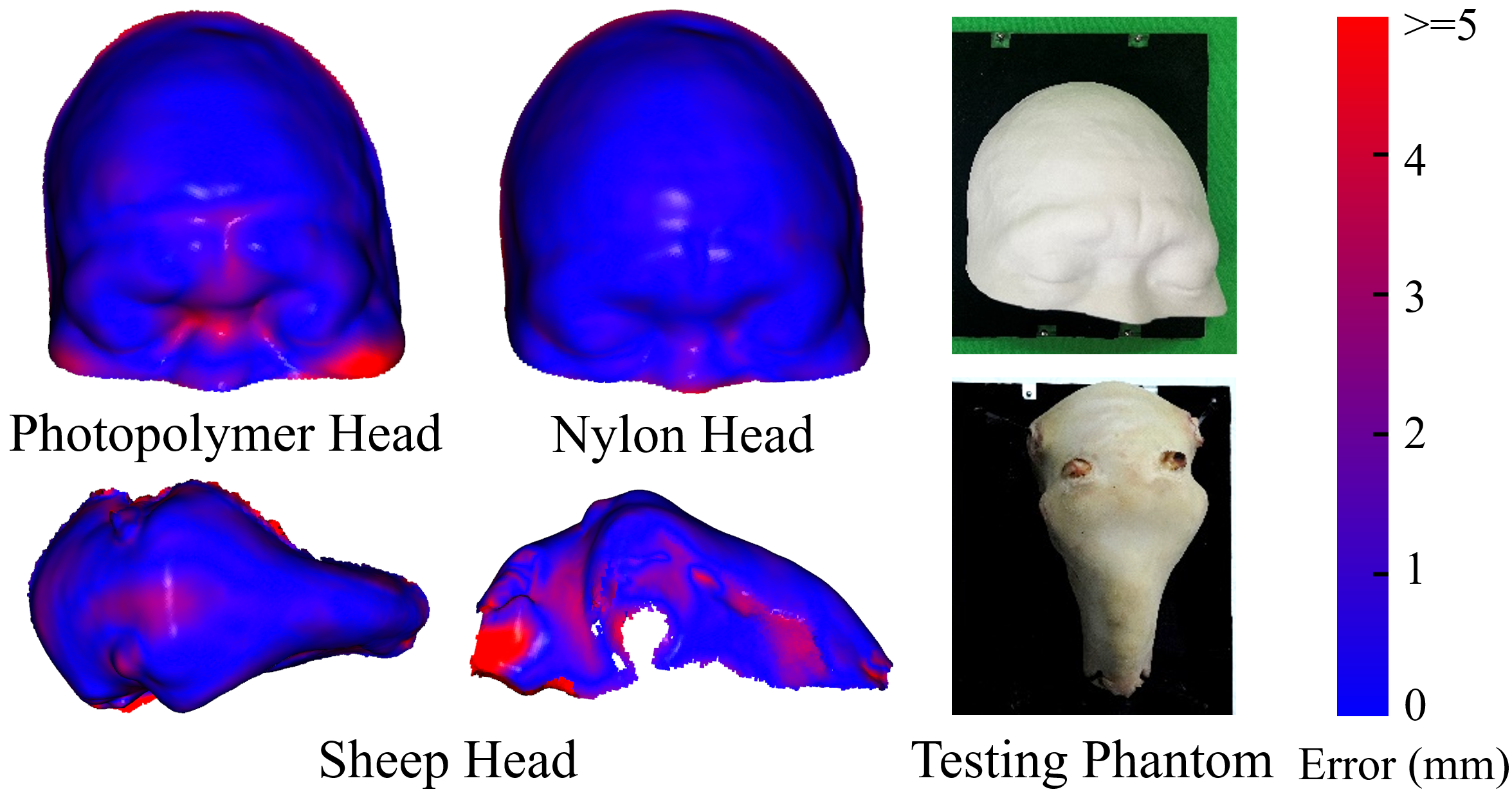}
    \caption{Error distributions of the reconstructed target surface on head phantoms of different materials and on sheep head surface, utilizing CT scanning as the ground truth.}
    \label{figure:ReconstructionErrorDistribution}
\end{figure}

This experiment was first conducted on head phantoms, 3D printed with both photopolymer and nylon, while a real sheep head was also included to better prove the performance in real situations (see Fig. \ref{figure:ReconstructionErrorDistribution}).
The quantitative results are recorded in Table. \ref{table:ResultReconstructionerror}. 
As the distributions of the results were non-normal for all the testing groups, the median value and IQR of the error distribution over all the points on the reconstructed surface were recorded. 
As a result, the median reconstruction errors were $0.56mm$ and $0.46mm$ for head phantoms, on photopolymer and nylon respectively.
The reconstruction accuracy of the sheep head was marginally larger than that of the head phantoms but still reached sub-millimetre accuracy ($0.79 mm$ median error, $1.06mm$ IQR).
Most of the high-error areas were located at the eye corners and the margins of the phantoms, which corresponded to areas of high depth error.
A similar situation was observed on the sheep's head.

\begin{figure}[htpb]
    \centering
    \includegraphics[width=1.0\columnwidth]{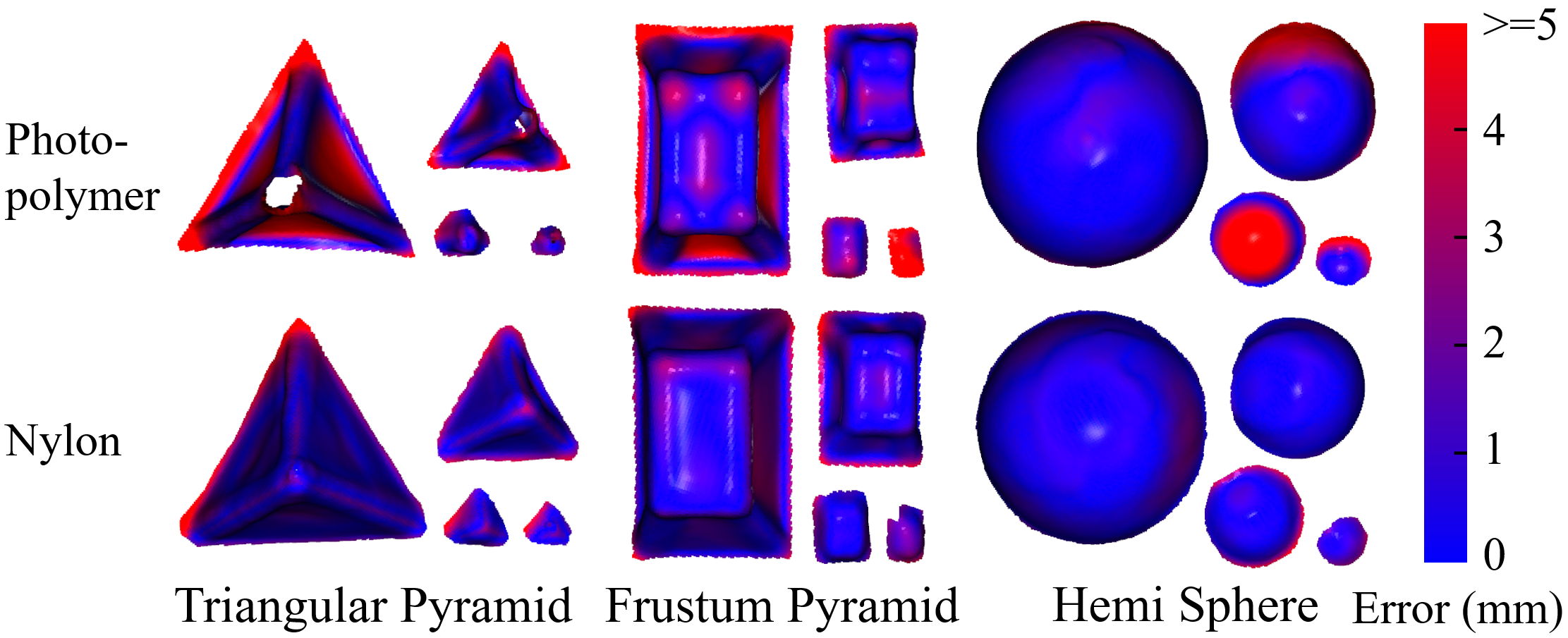}
    \caption{Error distributions of reconstructed target surface on models of different shapes, sizes and materials.}
    \label{figure:ReconstructionErrorDistributionAll}
\end{figure}

To better demonstrate the performance of the proposed surface reconstruction method and explore the limitations of such an approach, similar experiments were conducted on the models shown in the previous experiment, with two different materials, three different shapes and four different sizes.
The results are recorded in Fig. \ref{figure:ReconstructionErrorDistributionAll} and Table. \ref{table:ResultReconstructionerror}.

Regarding material, nylon demonstrated a more minor median reconstruction error than photopolymer across all the experiment groups despite the hemisphere (size 4).
Similar to the results on depth error, the reconstruction accuracy was high on the edge of the triangular pyramid models and frustum pyramid models, but more significant errors were revealed near these edges.
This may explain the error distributions on head phantoms and sheep head, that the surface near a strong convex edge may obtain lower reconstruction accuracy.
Meanwhile, significant errors were revealed at the corners of the models.
When comparing results on the same material, head phantoms consistently exhibited the least reconstruction error, while the errors on hemisphere were lower than those of triangular pyramid models and frustum pyramid models when the sizes of the models were large (size 1 and size 2), for both photopolymer and nylon.
Notably, the reconstruction errors were lower than the depth errors, in all the experiment groups, despite that the reconstruction error on the triangular pyramid (size 1) was $0.02mm$ higher than its depth error.
On average, the median reconstruction errors were $38.79\%$ and $56.03\%$ lower than the depth errors, for photopolymer and nylon respectively.

\begin{table}[htpb]
\caption{Median value and IQR of reconstruction error (mm) with depth correction.}
\centering
\begin{tabular}{ccccccccc}
\toprule
                     & \multicolumn{2}{c}{Size 1}                          & \multicolumn{2}{c}{Size 2}                  & \multicolumn{2}{c}{Size 3}                  & \multicolumn{2}{c}{Size 4}                  \\
                     & Med                      & IQR                      & Med                  & IQR                  & Med                  & IQR                  & Med                  & IQR                  \\ \cmidrule{2-9} 
                     & \multicolumn{8}{c}{Photopolymer}                                                                                                                                                              \\ 
H                    & 0.56                     & 0.61                     &                      &                      &                      &                      &                      &                      \\
TP                   & 2.32                     & 3.27                     & 1.31                 & 1.85                 & 1.46                 & 1.23                 & 1.98                 & 1.02                 \\
FP                   & 1.88                     & 2.92                     & 0.98                 & 1.22                 & 2.00                 & 1.33                 & 4.99                 & 1.50                 \\
HS                   & 0.64                     & 0.79                     & 0.97                 & 2.96                 & 3.37                 & 3.77                 & 1.13                 & 2.45                 \\ 
AVG                  & 1.61                     & 2.33                     & 1.09                 & 2.01                 & 2.28                 & 2.11                 & 2.70                 & 1.66                 \\ 
\cmidrule{2-9} 
                     & \multicolumn{8}{c}{Nylon}                                                                                                                                                                     \\ 
H                    & 0.46                     & 0.42                     &                      &                      &                      &                      &                      &                      \\
TP                   & 0.80                     & 0.82                     & 0.77                 & 0.95                 & 0.97                 & 1.19                 & 0.96                 & 1.11                 \\
FP                   & 1.16                     & 1.34                     & 0.87                 & 0.93                 & 0.89                 & 0.79                 & 1.84                 & 0.76                 \\
HS                   & 0.63                     & 0.48                     & 0.47                 & 0.28                 & 0.63                 & 0.78                 & 1.30                 & 0.89                 \\
AVG                  & 0.86                     & 0.88                     & 0.70                 & 0.72                 & 0.83                 & 0.92                 & 1.37                 & 0.92                 \\
\cmidrule{2-9} 
                     & \multicolumn{8}{c}{Sheep Head}                                                                                                                                                                 \\ 
\multicolumn{1}{l}{} & \multicolumn{1}{l}{0.79} & \multicolumn{1}{l}{1.06} & \multicolumn{1}{l}{} & \multicolumn{1}{l}{} & \multicolumn{1}{l}{} & \multicolumn{1}{l}{} & \multicolumn{1}{l}{} & \multicolumn{1}{l}{} \\ \bottomrule
\multicolumn{9}{p{220pt}}{H: Head Phantom; TP: Triangular Pyramid; FP: Frustum Pyramid; HS: Hemi-Sphere; AVG: Average value over TP, FP and HS.}
\end{tabular}
\label{table:ResultReconstructionerror}
\end{table}

\section{Use Cases}
\label{sec:UseCases}
To further prove the accuracy of in-situ guidance utilizing our proposed framework, a use case study was introduced to simulate EVD surgery.
The setup of this experiment is shown in Fig. \ref{fig:TeaserFigure} (c).
The testing structure comprises a 3D-printed head phantom in Nylon and two retro-reflective tools, rigidly fixed on an acrylic board to simulate the head frame.
Silicon gel was filled in the head phantom, while three holes were exposed to inject k-wires on the left, right and top of the head, respectively.
In addition, eight metal balls are fixed on the testing structure, which is visible in CT scanning, to enable the registration of pre-operative and post-operative images.
The structure was first virtualized with CT scanning, and the volume information was utilized to plan multiple trajectories targeting the ventricle.
For each hole on the head phantom, three trajectories were planned, leading to 9 trajectories in total.
Despite the injection trajectories, the intersection points of the trajectories and the surface of the silicon gel were also extracted to indicate the entry points.  
Before the conduction of the experiment, three retro-reflective markers $\Phi = 10mm$ were stuck on the phantom, while 1000 frames of sensor data were utilized to acquire the parameters for the depth distortion model of the head phantom. 

\begin{figure}[b]
    \centering
        \includegraphics[width=1.0\columnwidth,page=1]{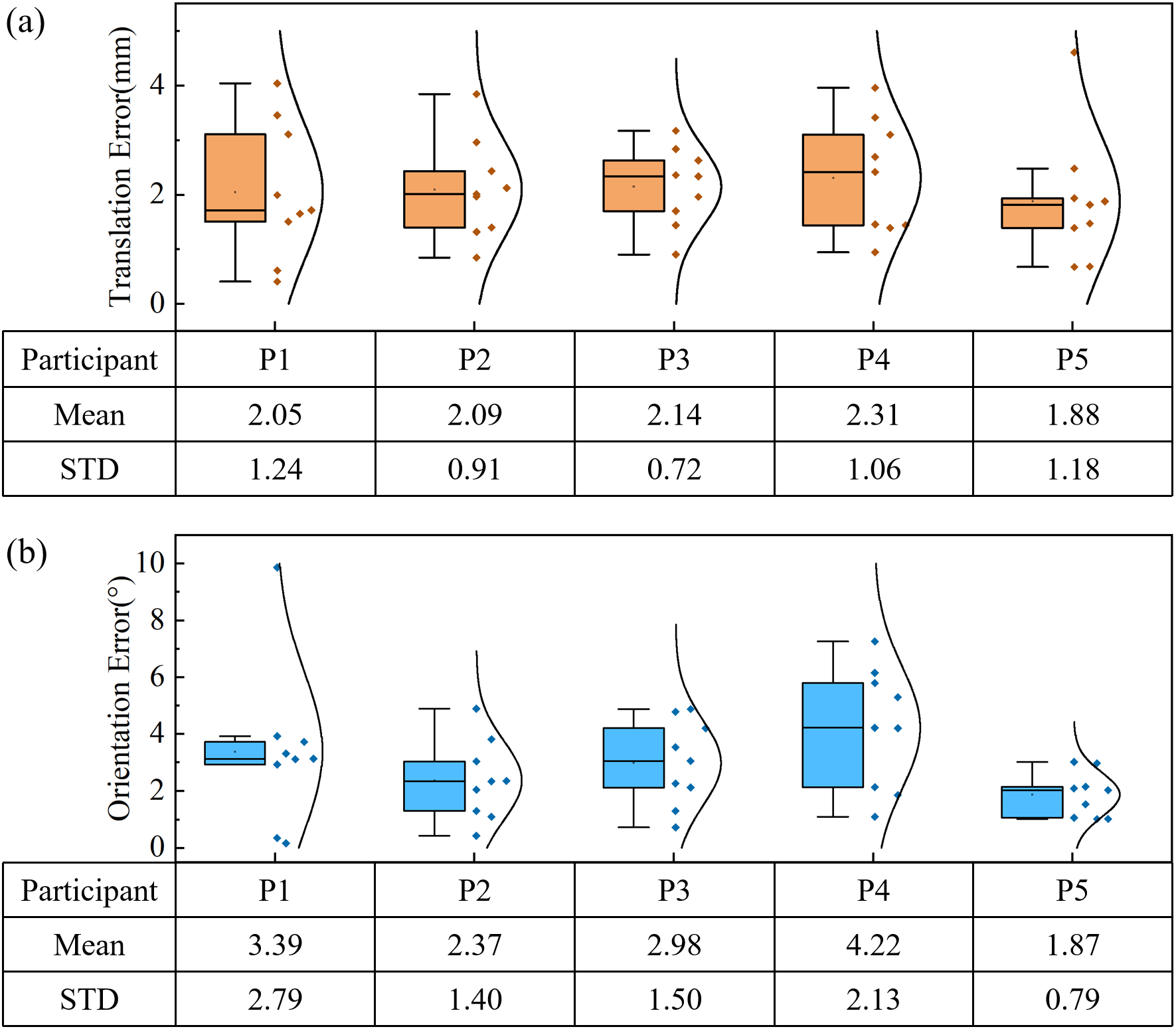}
    \caption{Error in translation and rotation for simulated EVD injections using proposed head tracking framework for visual guidance. Five surgeons (P1-P5) were included in this study.}
    \label{fig:ResultUseCase}
\end{figure}

In total, five surgeons were included in this study, and each performed nine k-wire injections.
For each group of experiments, an eye calibration was first performed.
After that, 1000 frames of ToF sensor data were recorded to construct the shape of the retro-reflective tool and reconstruct the head surface.
The two retro-reflective tools fixed on both sides of the head phantom were regarded as one rigid tool.
The head surface extracted from the pre-operative image was then registered with the reconstructed surface to obtain transformation from planned trajectories to the retro-reflective tools.
The entry point was displayed in-situ as a sphere for each trajectory, while the trajectory was displayed as a line.
Before the formal experiment, one random trajectory was generated for each hole on the phantom for surgeons to get familiar with the system.
After the warm-up procedure, the surgeons were asked to perform the injections according to the planned trajectories.
To better prove the accuracy of the in-situ display, no visual guidance for the drill and the injected k-wire was provided during the experiment.
To measure the performance of the simulated EVD injections, the pre-operative phantom was scanned with CT and registered with pre-operative images utilizing metal balls fixed on the phantom.
The rotation and shift of entry points between real trajectories and the planned ones were calculated for evaluation.
In general, the five participants included in this study reported an injection accuracy of $2.09\pm 0.16 mm$ in entry points, and $2.97\pm 0.91^\circ$ for orientations (see Fig. \ref{fig:ResultUseCase}).
These results showcased a comparable accuracy to other surgical navigation systems performing tracking with AR-HMD \cite{liebmann2019pedicle, spirig2021augmented}, and is higher than those implementing retro-reflective tools for tracking and did not include registration of pre-operative images \cite{martin2023sttar}.

\section{Discussions}
\label{sec:Discussions}
This work introduces a framework utilizing the built-in ToF depth camera on commercially available AR-HMDs, to enable visual guidance in EVD surgery without attaching tracking targets to the patient's head invasively. 
The pilot studies conducted in the paper revealed a systematic depth error in ToF sensors on AR-HMDs, which varied greatly depending on materials, target depths, and surface orientations.
The error was tested to be $7.580\pm 1.488$ on the human head surface in an experiment including 10 participants.
These results suggested that depth data from the ToF sensor needs to be corrected patient-specifically for accurate surgical guidance.

The proposed framework is shown in Fig. \ref{fig:SystemStructure}, while Fig. \ref{fig:TeaserFigure} (a) showcases a setup for the proposed method.
A patient-specific ToF sensor correction method is first included for accurate surface information.
During the surgery, retro-reflective markers are attached to the head frames fixing the patient's head.
Corrected surface information provided by ToF sensor is utilized to bridge pre-operative images and the retro-reflective tool, where surface reconstruction is adopted to reduce noise and improve accuracy.
The retro-reflective tool is finally used for in-situ guidance with high efficiency.
A system combining AR-HMD and 3D Slicer has been constructed to streamline the procedure.

Results regarding depth correction effectiveness before and after depth value correction indicated that the proposed method could significantly reduce depth error for most models and setups, as demonstrated in Table. \ref{table:ResultDeptherror}.
As the pipeline incorporated a material-specific depth error correction procedure, considerable error reductions were revealed for different materials, even when they performed widely varying depth errors in pilot studies (see Fig. \ref{fig:PrintingMaterialError} (b), \ref{fig:PrintingMaterialError} (c)).
On the head phantoms specifically, only $0.89mm$ depth error remained for photopolymer after correction ($93.35\%$ reduction), which was $1.08mm$ for nylon ($85.87\%$ reduction).
Results on three standard models of four different sizes indicated that the average median depth error across different models decreased after depth undistortion among all the sizes and both materials.
However, the reduction ratio decreased as the models became smaller, which may limit the performance of providing depth information for small objects.
An inappropriate estimation of the surface norm, when the models are small, may potentially contribute to this reduction.
Moreover, the depth value on the acrylic board used to hold the target objects may potentially influence the accuracy at the edge of the models, which might account for the relatively large error at the margin of the models surface (see Fig. \ref{figure:ReconstructionErrorDistribution}).
Though depth error decreased statistically when models became small before undistortion, this should not imply that the depth camera would be more accurate for smaller target objects.
Instead, this tendency was caused by mismatches between points on undistorted surfaces and those on ground truth when large offsets exist during error calculation.
Furthermore, our results for surface reconstruction proved that the proposed method could reconstruct the target surface with sub-millimetre accuracy.
The reconstruction accuracy on two head phantoms were $0.56mm$ and $0.46mm$, while $0.79 mm$ accuracy was revealed for a real sheep head.
When comparing depth errors and reconstruction errors for the same objects, the reconstructed surface provided lower error in all the experiment groups, while only one group of study provided a reconstruction error $0.02mm$ higher than the depth error.
This proved the necessity to use reconstruction during pre-operative registration.
Moreover, similar characteristics of error distributions were found on depth data error after correction and reconstruction accuracy.
The errors were smaller in smoother areas and had larger radii of curvatures.
Inversely, more significant errors were found near the strong edges or concavities, such as the inner corners of the eyes on the head phantoms, or the edges on the triangular pyramid models.
Adding that the performance of the depth correction was comparatively limited on small objects, it may indicate a limitation of the proposed method to provide surface information for complex, small tissues.

A use case study simulating drainage tube insertion was finally conducted in our work to prove the applicability of the proposed framework to provide visual guidance during EVD surgery.
The factors influencing the results of this experiment comprised the errors from both pre-operative registration and intra-operative tracking.
The surgical tool was not virtualized in this experiment to avoid the relative spatial information provided by two virtual objects.
Five surgeons were included in this study, and each of them performed 9 k-wire injections following virtual guidance.
In the end, $2.09\pm 0.16mm$ accuracy in entry points and $2.97\pm0.91^\circ$ in orientation were revealed.

Compared with previous works, the proposed framework integrates retro-reflective tool tracking and surface information to realize accurate EVD guidance, while the tracking target is not rigidly fixed on the patient's head to avoid invasion.
While depth data errors in the ToF sensor on AR devices have been revealed in existing studies, to the best of our knowledge, this study is the first to correct this error, significantly improving the potential accuracy for in-situ guidance.
In existing surgical guidance studies utilizing retro-reflective targets for tracking, Alejandro et al.'s study \cite{martin2023sttar}, reported $2.79\pm 1.52 mm$ entry point accuracy and $4.56\pm2.49^\circ$ orientation accuracy in the use case study including 7 surgeons for simulating k-wire injections.
Meanwhile, Van et al. reported $11.9\pm 4.5mm$ target point accuracy in EVD simulation \cite{van2021effect}.
Comparatively, our study extends the procedure to include both registration between the patient and the tracking target, and intra-operative tracking, using solely on-device ToF depth sensor and still reached comparable results to previous studies ($2.09\pm 0.16mm$ entry point accuracy, $2.97\pm0.91^\circ$ orientation accuracy).

Moreover, although this work primarily aims to provide accurate visual guidance for EVD surgery, the methods developed here have broader potential applications, including remote consultants, surgical collaboration, etc.
While there have been several studies proposed for spatial reconstruction using implemented sensor resources in AR head-mounted devices \cite{jung2021model, wu20203d}, to our knowledge, this is the first instance of a dense surface reconstruction method with sub-millimetre accuracy with only a ToF sensor on AR devices.
This achievement can be attributed to including the retro-reflective targets for accurate pose estimation and the depth sensor correction method, providing more accurate surface information.
We believe similar approaches would not only be applicable to AHAT depth sensor mode from HoloLens 2, but also to other devices or long throw depth mode from HoloLens, enabling dense, accurate surface reconstruction of various scales.

While the proposed pipeline offers a clear and simple approach to enable accurate EVD surgical guidance, certain limitation exists with this framework.
The proposed depth sensor correction method assumes that the depth error across the patient's head remains uniform.
However, different areas on the head surface may perform varying depth errors, leading to distortions in the reconstructed head surface.
Moreover, distortion is observed on the reconstructed faces near the strong edges of the target objects, which cannot be corrected utilizing the proposed correction model.
Consequently, reconstruction accuracy may be lower in areas with complex shapes and pronounced edges.
If further applications demand higher reconstruction accuracy, the issue of depth correction on non-uniform, complex surfaces must be addressed.
Alternatively, the inclusion of stereo visible light cameras on commercial AR devices in the reconstruction pipeline could potentially improve the performances, particularly in areas with sharp features.

\section{Conclusion}
\label{sec:Conclussion}
This paper proposes a framework for accurate surgical guidance during EVD surgery with high efficiency and no extra invasion.
The proposed method takes advantage of both retro-reflective tool tracking and point cloud information to enable pre-operative registration and real-time guidance using solely ToF depth sensor integrated into commercially available AR-HMD.
The ToF sensor depth correction method proposed in this study showed a capability to reduce depth error by over $85\%$ on head phantoms in different materials.
The proposed surface reconstruction method achieved sub-millimetre accuracy on head phantoms.
A real sheep head included in the experiments showed $0.79 mm$ reconstruction accuracy.
In the end, a use case study simulating drainage tube insertion was conducted to demonstrate the feasibility of utilizing the proposed system for EVD guidance.
Results of this experiment showcased operation accuracy of $2.09\pm 0.16mm$ in entry point and $2.97\pm 0.91^\circ$ in orientation.


\bibliographystyle{IEEEtran}
\bibliography{Bibliography}
\end{document}